\documentclass[letterpaper, 10 pt, conference]{ieeeconf} 

\IEEEoverridecommandlockouts                             
\usepackage{amsmath} 
\usepackage{amssymb}  
\usepackage{textcomp}

\newtheorem{myrem}{Remark}

\newtheorem{mypro}{Proposition}

\usepackage{graphicx}
\usepackage{textcomp}
\usepackage{float}
\newcommand*{\QEDA}{\hfill\ensuremath{\blacksquare}} 
\newenvironment{prooflemma}
{{\indent\itshape Proof:}\,}{\hfill$\QEDA$}

\usepackage[linesnumbered,boxed,ruled,commentsnumbered]{algorithm2e}
\SetKwRepeat{Do}{do}{while}%
\usepackage{color}

\title{\LARGE \bf
Routing Guidance for Emerging Transportation Systems \vspace{1.5pt} 

with Improved Dynamic Trip Equity
}

\author{Ting Bai, Anni Li, Gehui Xu, Christos G. Cassandras,~\IEEEmembership{Life Fellow, IEEE}, and Andreas A. Malikopoulos
\thanks{This research was supported in part by NSF under Grants CNS-2401007, CMMI-2348381, IIS-2415478, and in part by MathWorks.}\vspace{1.5pt}
\thanks{T. Bai and A. A. Malikopoulos are with the Information and Decision Science Lab, School of Civil $\&$ Environmental Engineering, Cornell University, Ithaca, New York, U.S.A. E-mails: \{{\tt\small tingbai, amaliko\}@cornell.edu}}
\thanks{A. Li and C. G. Cassandras are with the Division of Systems Engineering and Center for Information and Systems Engineering, Boston University, Brookline, Massachusetts, U.S.A. E-mails: \{{\tt\small anlianni, cgc\}@bu.edu}}
\thanks{G. Xu is with the Department of Electrical and Electronic Engineering, Imperial College London, SW7 2AZ London, U.K. E-mail: {\tt\small g.xu@imperial.ac.uk}}}
\begin{document}

\maketitle
\thispagestyle{empty}
\pagestyle{empty}

\begin{abstract}
This paper presents a dynamic routing guidance system that optimizes route recommendations for individual vehicles in an emerging transportation system while enhancing travelers' trip equity. We develop a framework to quantify trip quality and equity in dynamic travel environments, providing new insights into how routing guidance influences equity in road transportation. Our approach enables real-time routing by incorporating both monitored and anticipated traffic congestion. We provide conditions that ensure perfect trip equity for all travelers in a free-flow network. Simulation studies on 1,000 vehicles traversing an urban road network in Boston demonstrate that our method improves trip equity by approximately 11.4\% compared to the shortest-route strategy. In addition, the results reveal that our approach redistributes travel costs across vehicle types through route optimization, contributing to a more equitable transportation system.
\end{abstract}

\section{Introduction}\label{Section I}
Transportation is a fundamental pillar of modern society, enabling access to essential daily services and activities, such as education, employment, healthcare, and social interactions. Recent advancements in transportation technologies, including shared mobility~\cite{shaheen2016mobility}, electric vehicles~\cite{Bai2025distributed,10147895}, and automated driving systems~\cite{woo2021understanding}, are transforming travel patterns. Specifically, shared mobility reduces reliance on private vehicle ownership, electric vehicles lower travel costs while mitigating greenhouse gas emissions, and automated driving enhances road safety and traffic efficiency. These innovations hold great potential to shape a more efficient, cost-effective, and sustainable transportation system.  

Equity has been an important principle in transportation to promote fairness and inclusivity~\cite{yan2020fairness}. It advocates for the just distribution of transportation resources, services, and infrastructure, ensuring that all individuals, regardless of their socioeconomic status, race, gender, or geographic location, have equal access to safe, affordable, and reliable mobility options. Transportation solutions lacking access equity can lead to unintended negative consequences~\cite{bang2024emergingequity}. As noted in~\cite{carleton2018comparative}, low-income populations, who often rely on public transit, face substantial barriers when affordable and reliable transportation options are limited. This inequity restricts their access to essential services and opportunities, and reinforces cycles of poverty and social exclusion. In addition, disparities in transportation access can drive over-reliance on private vehicles, exacerbating traffic congestion, increasing travel costs, and contributing to environmental degradation. 

Advancing equity in emerging transportation systems requires developing routing guidance systems that address the diverse needs of travelers, such as availability and affordability, while prioritizing equity. Different from conventional routing guidance services~\cite{toth2014vehicle}, which focus on optimizing traffic efficiency through metrics such as shortest distance, minimal time, or cost, equity-based routing guidance systems integrate fairness into existing frameworks, delivering socially optimal solutions for travelers despite their disparities. 

Existing routing guidance systems fall into two categories: \textit{static} and \textit{dynamic} systems~\cite{jeihani2022investigating}. Static routing guidance systems (S-RGSs) rely on macroscopic approaches to generate flow-level route recommendations, offering uniform guidance to all vehicles in the same flow. These systems are less accurate in computing optimal routes as they fail to incorporate real-time traffic information, such as congestion or incidents, for updating routes. Moreover, since all vehicles in the same flow are treated the same, static routing strategies may shift congestion from one point to another rather than effectively resolve it~\cite{liang2014route}. In contrast, dynamic routing guidance systems (D-RGSs) leverage real-time traffic data to continuously update route recommendations, resulting in more accurate and adaptive routes~\cite{fu2001adaptive,deflorio2003evaluation}. While dynamic systems offer improved accuracy, studies on developing effective D-RGSs are still scarce, particularly in the context of enhancing mobility equity.

In this paper, we present a D-RGS designed to optimize the route recommendations for individual vehicles in an emerging transportation system with enhanced trip equity. Unlike existing D-RGSs, our approach integrates equity considerations into the decision-making process, ensuring that the benefits of transportation advancements are distributed fairly across diverse populations. The main contributions are as follows: (i) We present a novel framework to measure the quality and equity of trips in a \textit{dynamic} emerging transportation system through the notions of the dynamic trip index and dynamic trip equity. (ii) Based on the proposed metrics, we derive conditions for achieving perfect trip equity in free-flow networks, which provide the upper bound for trip equity attainable through route optimization. (iii) We propose a D-RGS that incorporates both monitored and estimated traffic congestion in real-time route optimization and allows for trip equity improvement. Finally, simulation studies on $1,000$ vehicles traversing an urban road network in Boston demonstrate the effectiveness of the proposed framework. 

\section{Problem Formulation}\label{Section II}
\subsection{Road Network}
We model the road network as a directed graph $\mathcal{G}(\mathcal{V},\mathcal{E})$, where $\mathcal{V}$ is the set of nodes and $\mathcal{E}\!\subset\!{\mathcal{V}\!\times\!{\mathcal{V}}}$ is the set of edges. Each node $v\!\in\!\mathcal{V}$ is uniquely indexed, representing a key intersection or an origin/destination node in the network, while each edge $(v,v^{\prime})\!\in\!\mathcal{E}$ represents a road segment connecting a pair of nodes. The weight of each edge, denoted by $\tau(v,v^{\prime})\!\in\!\mathbb{R}_{+}$, denotes the travel time to traverse the edge, where $\mathbb{R}_{+}$ is the set of positive real numbers. We denote the free-flow travel time on an edge $(v,v^{\prime})$ by $\tau_0(v,v^{\prime})$.

Consider an emerging transportation system consisting of diverse transport modes, such as private vehicles and autonomous vehicles. At any time~$t$, let $N(t)\!\in\!{\mathbb{N}}$ denote the number of vehicles \textit{actively} traveling in the network. By ``actively," we refer to the vehicle taking at least one traveler en route to a designated destination. For vehicle $i\!\in\!\mathcal{N}(t)\!:=\!\{1,2,\dots,N(t)\}$ traveling from a node $v_1$ to node $v_n$, a \textit{feasible route} between $v_1$ and $v_n$ is defined as
\begin{align}
r_{1,n}\!:=\!\big\{(v_1,v_2),(v_2,v_3),\dots,(v_{n-1},v_n)\big\},\label{Eq.1}
\end{align}
where $(v_k,v_{k+1})\in\mathcal{E}$, $k\!=\!1,\dots,n\!-\!1$. The set of all feasible routes from $v_1$ to $v_n$ is denoted by $\mathcal{R}_{1,n}$. In the routing problem, each vehicle $i$ is associated with a trip defined by its origin-destination (OD) pair, denoted as $(o_i,d_i)$. Namely, $v_1$ in \eqref{Eq.1} is initialized as $o_i$, and $v_n$ is fixed as $d_i$. The nodes $\{v_k\}$, $k\!=\!1,\dots,n\!-\!1$, in the feasible route are defined as a sequence of Decision-Making Points (DMPs). At each DMP $v_k$, the vehicle's route from $v_k$ to $d_i$ is optimized based on its location and real-time traffic conditions. This iterative process enables routing solutions to dynamically adapt to vehicle movements and evolving network dynamics.
\vspace{-3pt}

\subsection{Vehicle Dynamics}
To describe vehicle movements between DMPs, we introduce the following dynamic model. For each vehicle $i\!\in\!\mathcal{N}(t)$, let $a_k^i$ and $a_{k+1}^i$ be its arrival times at two consecutive DMPs $v_k$ and $v_{k+1}$, respectively. The dynamics of vehicle $i$ is depicted as 
\begin{align}
a^i_{k+1}=a^i_k+\tau_{k,k+1}^i,\label{Eq.2}
\end{align}
where $\tau_{k,k+1}^i$ is the travel time of vehicle $i$ to traverse edge $(v_k,v_{k+1})$, which can be modeled using the widely adopted Bureau of Public Road (BPR) function~\cite{united1964traffic}. The BPR model captures the impact of traffic congestion by describing how travel time increases as traffic demand approaches or exceeds road capacity. Specifically, $\tau_{k,k+1}^i$ is of the form
\begin{align}
\tau_{k,k+1}^i=\tau_{k,k+1}^0\Bigg(\!1\!+\!\alpha\!\left(\frac{f_{k,k+1}^i}{c_{k,k+1}}\right)^{\beta}\!\Bigg),\label{Eq.3}
\end{align}
where $\tau_{k,k+1}^0$ denotes the free-flow travel time on $(v_k,v_{k+1})$, $f_{k,k+1}^i$ denotes the traffic flow on the edge when vehicle~$i$ passes through, and $c_{k,k+1}$ is the road capacity. The parameters $\alpha$ and $\beta$ characterize the sensitivity of travel time to traffic flow, with typical values of $\alpha\!=\!0.15$ and $\beta\!=\!4$. 

In a dynamic system, $f_{k,k+1}^i$ varies over time, depending on the number of vehicles traveling on $(v_k,v_{k+1})$ within the same time interval. Next, we present appropriate monitoring and estimation approaches to characterize the traffic flow.

\subsection{Traffic Flow Monitoring $\&$ Estimation}
At a DMP $v_k$, the traffic flow on each edge forming the potential feasible routes from $v_k$ to $d_i$ needs to be monitored or estimated for optimal routing. To this end, we compute the traffic flow in two phases: (i) monitoring the traffic flow on \textit{adjacent edges}, which are directly adjacent to $v_k$, and (ii) estimating the traffic flow on \textit{future edges}, which are edges further along the feasible routes to the destination. 

\subsubsection{Flow Monitoring for Adjacent Edges} Let $(v_k,v_{k+1})$ be an adjacent edge along the feasible route $r_{k,d_i}$ from $v_k$ to $d_i$, where $v_{k+1}\!\in\!{\mathcal{V}_{k+1}}$ may not be unique with $\mathcal{V}_{k+1}$ denoting the set of alternative nodes for $v_{k+1}$. Given the arrival time $a_k^i$, a time window for monitoring the traffic flow on $(v_k,v_{k+1})$ is defined as
\begin{align}
\mathcal{M}\big(a_k^i,\Delta{t}\big)\!:=\!\big[a_k^i\!-\!\Delta{t}, a_k^i\!+\!\Delta{t}\big],\label{Eq.4}
\end{align}
where $\Delta{t}\!>\!{0}$ denotes the size of the monitoring window. Accounting for other vehicles $j\!\in\!\mathcal{N}(t)\!\setminus\!\{i\}$ entering the same edge within this window, we define the indicator function 
\begin{align}
    \!\!\mathbf{1}_{a_{k,k+1}^j}\!:=\!\begin{cases}
	1,& \text{if~ $a_{k,k+1}^j\!\in\!{\mathcal{M}\big(a_k^i,\Delta{t}\big)}$,} \\
	0,& \text{otherwise,}
	\end{cases}\label{Eq.5}
\end{align}
where $a_{k,k+1}^j$ denotes the time when vehicle $j$ arrives at $v_k$ and enters $(v_k,v_{k+1})$, satisfying $a_{k,k+1}^j\!=\!a_k^j$. As such, the traffic flow on each adjacent edge upon vehicle $i$'s arrival can be monitored and computed by
 \begin{align}
 \tilde{f}_{k,k+1}^i=\frac{\sum_{j\in\mathcal{N}(t)\setminus\{i\}}{\!\mathbf{1}_{a_{k,k+1}^j}\!+\!1}}{2\Delta{t}},\label{Eq.6}
 \end{align}
including vehicle $i$ itself. The total length of the monitoring time window is $2\Delta{t}$, as defined in \eqref{Eq.4}.

\subsubsection{Flow Estimation for Future Edges} For future edges $(v_s,v_{s+1})\!\in\!{r_{k,d_i}}$, $s\!\neq\!{k}$, the traffic flow at arrival time $a_s^i$ cannot be directly observed at time $a_k^i$. However, by leveraging the route plans of other vehicles $j\!\in\!\mathcal{N}(t)\!\setminus\!\{i\}$, the traffic flow can be anticipated, enabling congestion estimation and route optimization for vehicle $i$ at each of its DMPs.

To proceed, let $r_{k^{\prime},d_j}^*(a_k^i)$ be the optimal route plan of vehicle $j\!\in\!\mathcal{N}(t)\!\setminus\!\{i\}$ at time $a_k^i$ for completing its remaining trip, where $k^{\prime}$ indicates that the route plan was made at DMP $v_{k^{\prime}}$. For each future edge $(v_s,v_{s+1})\!\in\!{r_{k,d_i}}$, $s\!\neq\!{k}$, the arrival time of vehicle $i$ at $v_s$ can be estimated by \eqref{Eq.2}, i.e., 
\begin{align}
\hat{a}_s^i=\begin{cases}a_k^i+\tilde{\tau}_{k,k+1}^i,& \text{if $s\!=\!k\!+\!1$,}\\
\hat{a}_{s-1}^i+\hat{\tau}_{s-1,s}^i,&\text{otherwise,}
\end{cases}\label{Eq.7}
\end{align}
where $\tilde{\tau}_{k,k+1}^i$ representing the time for vehicle $i$ to traverse edge $(v_k,v_{k+1})$ is computed by integrating the monitored traffic flow $\tilde{f}_{k,k+1}^i$ in \eqref{Eq.3}. Additionally, $\hat{\tau}_{s-1,s}^i$ in~\eqref{Eq.7} denotes the estimated travel time for vehicle $i$ to traverse the edge $(v_{s-1},v_s)$. Using the estimated arrival time $\hat{a}_s^i$, the traffic flow on the future edge $(v_s,v_{s+1})$ can be estimated by
 \begin{align}
 \hat{f}_{s,s+1}^i\!=\!\frac{\sum_{(v_s,v_{s+1})\in{r}_{k^{\prime},d_j}^*\!(a_k^i),j\in\mathcal{N}(t)\setminus\{i\}}\!{\mathbf{1}_{\hat{a}_{s,s+1}^j}}\!\!+\!1}{2\Delta{t}},\label{Eq.8}
 \end{align}
where $\hat{a}_{s,s+1}^j$ denotes the estimated time when vehicle~$j$ arrives at $v_s$ and enters $(v_s,v_{s+1})$, which is determined by the dynamic model of vehicle $j$ considering nominal travel times along $r_{k^{\prime},d_j}^*(a_k^i)$. Similar to \eqref{Eq.5}, the indicator function $\mathbf{1}_{\hat{a}_{s,s+1}^j}\!$ in \eqref{Eq.8} is defined as
\begin{align}
    \mathbf{1}_{\hat{a}_{s,s+1}^j}\!:=\!\begin{cases}
	1,& \text{if~ $\hat{a}_{s,s+1}^j\!\in\!{\mathcal{M}\big(\hat{a}_s^i,\Delta{t}\big)}$,} \\
	0,& \text{otherwise.}
	\end{cases}\label{Eq.9}
\end{align}
Integrating $\hat{f}_{s,s+1}^i$ into \eqref{Eq.3}, the estimated travel time $\hat{\tau}_{s,s+1}^i$ and arrival time $\hat{a}_{s+1}^i$ can be obtained accordingly. Note that for each future edge $(v_s,v_{s+1})\!\in\!{r_{k,d_i}}$, $s\!\neq\!{k}$, the estimated traffic flow $\hat{f}_{s,s+1}^i$, travel time $\hat{\tau}_{s,s+1}^i$, and the resulting estimated arrival time $\hat{a}_{s+1}^i$ are computed iteratively.
\vspace{-8pt}

\subsection{Routing Problem}
In our exposition, we consider an emerging transportation system integrating (1) \textit{private vehicles}, (2) \textit{autonomous vehicles}, and (3) \textit{ride-hailing vehicles}. The sets of vehicles in each category are denoted by $\mathcal{N}_p(t)$, $\mathcal{N}_a(t)$, $\mathcal{N}_h(t)$, with $\mathcal{N}(t)\!=\!\mathcal{N}_p(t)\!\cup\!{\mathcal{N}_a(t)}\!\cup\!{\mathcal{N}_h(t)}$. For simplicity, we assume travelers taking ride-hailing vehicles share the same OD pairs and split costs equally, while private and autonomous vehicles serve only one traveler at a time. Owing to automation technology, autonomous vehicles have a lower per-mile cost than that of private vehicles. 

While individuals select their travel modes based on factors such as income, accessibility, and reliability, the quality of their trips should remain comparable to ensure equitable access to transportation resources and services. This paper aims to develop a D-RGS guide each vehicle $i\!\in\!\mathcal{N}(t)$ in selecting an optimal route $r_{k,d_i}^*\!\in\!\mathcal{R}_{k,d_i}$ at each DMP $v_k$ while enhancing trip equity across all travelers. 

\section{Dynamic Trip Index and Trip Equity}\label{Section III}
This section presents a framework to evaluate the quality of each trip and the equity of all trips in a dynamic transportation system. The framework is built upon two key concepts: the \textit{dynamic trip index} and \textit{dynamic trip equity} that we introduce next.

\subsection{Dynamic Trip Index}
The trip index quantifies the quality of a trip in terms of its efficiency (i.e., travel time), cost, and convenience, which are key factors in the evaluation of transport quality. In a dynamic transportation system, the trip index is influenced by factors such as the travel mode selected, the vehicle's routing, and the schedules and routes of other vehicles. For any traveler taking a vehicle $i\!\in\!\mathcal{N}(t)$ for the trip $(o_i,d_i)$, the \textit{Dynamic Trip indeX} (DTX) evaluated at time $t$ is defined as 
\begin{align}
DTX_i(t):=\xi_1\frac{\tau_{\min}^i}{\tau_i(t)}+\xi_2\frac{\phi_{\min}^i}{\phi_i(t)}+\xi_3\frac{q_{\min}^i}{q_i},\label{Eq.10}
\end{align}
where the three terms represent the evaluations of efficiency, cost, and convenience of the trip, respectively. Specifically, $\xi_1,\xi_2,\xi_3\!\in\!\mathbb{R}_{+}$ are positive constants serving as weighting coefficients to balance the contribution of each factor in $DTX_i(t)$. These parameters satisfy the convex combination condition $\xi_1\!+\!\xi_2\!+\!\xi_3\!=\!1$, ensuring that the weights are normalized and appropriately distributed across the evaluated factors. In \eqref{Eq.10}, $\tau_{\min}^i$, $\phi_{\min}^i$, and $q_{\min}^i$ denote the minimum values of travel time, cost, and inconvenience achievable for completing the trip $(o_i,d_i)$ among all alternative transportation modes under ideal traffic conditions (i.e., without congestion), which are computed by
\begin{subequations}\label{Eq.11}
\begin{align}
\tau_{\min}^i&\in\arg\min_{r\in\mathcal{R}_{o_i,d_i}}\!\sum_{(v_{\ell},v_{\ell+1})\in{r}}\!\!\!\!\!\!\tau^0_{\ell,\ell+1},\label{Eq.11a}\\
\phi_{\min}^i&=\epsilon_{\min}^i\tau_{\min}^i,\label{Eq.11b}\\
q_{\min}^i&=\frac{T_{w,\min}^i}{T_{d,\max}^i},\label{Eq.11c}
\end{align}
\end{subequations}
where, as defined earlier, $\mathcal{R}_{o_i,d_i}$ is the set of all feasible routes from $o_i$ to $d_i$ and $\epsilon_{\min}^i\!:=\!\min\big\{\epsilon_p,\epsilon_a,\epsilon_h\big\}$ represents the minimum transportation cost per traveler per unit of travel time, with $\epsilon_p$, $\epsilon_a$, $\epsilon_h$ denoting the cost for private, autonomous, and ride-hailing vehicles, respectively. In \eqref{Eq.11c}, the best convenience of the trip (i.e., minimum inconvenience) is denoted by $q_{\min}^i$, where
\begin{subequations}\label{Eq.12}
\begin{align}
T_{w,\min}^i\!&=\min\big\{T_{w,p},T_{w,a},T_{w,h}\big\},\label{Eq.12a}\\
T_{d,\max}^i\!&=\max\big\{T_{d,p},T_{d,a},T_{d,h}\big\}.\label{Eq.12b}
\end{align}
\end{subequations}
We use $T_{w,p}$, $T_{w,a}$, $T_{w,h}$ to denote the average waiting times to access a private, autonomous, and ride-hailing vehicle, respectively. Here, $T_{w,p}$ accounts for the additional time required to park and retrieve a private vehicle, $T_{w,a}$ captures the potential delays due to scheduling and operational constraints of an autonomous vehicle, and $T_{w,h}$ reflects the average waiting time to dispatch a ride-hailing vehicle. In \eqref{Eq.12b}, $T_{d,p}$, $T_{d,a}$, $T_{d,h}$ denote the feasible departure time windows (in hours per day) for each vehicle type. 

The dynamic values of trip time, cost, and convenience are denoted by $\tau_i(t)$, $\phi_i(t)$, and $q_i$ in \eqref{Eq.10}, respectively. Let $r_i^*(t)$ be the optimal route plan for vehicle $i$ at time $t$ from its origin to the destination, which can be represented as 
\begin{align}
r_i^*(t)=r_{o_i,k}(t)\cup{r_{k,d_i}^*\!(t)},\label{Eq.13}
\end{align}
where $t\!\in\![a_k^i,a_{k+1}^i)$ and $r_{o_i,k}(t)$ denotes the experienced route of vehicle $i$ at time $t$. The optimal route planning for completing the rest of the trip is denoted as $r_{k,d_i}^*(t)$, which remains unchanged until the vehicle reaches its next DMP $v_{k+1}$. In light of \eqref{Eq.13}, we have
\begin{subequations}\label{Eq.14}
\begin{align}
\!\!\!\!\tau_i(t)&=\!\!\!\sum_{(v_{\ell},v_{\ell+1})\in{r_{o_i,k}(t)}}\!\!\!\!\!\!\!\!\!\!\!\!\!\tau_{\ell,\ell+1}^i\!+\tilde{\tau}_{k,k+1}^i+\!\!\!\!\!\!\!\!\!\sum_{(v_s,v_{s+1})\in{r_{k,d_i}^*\!(t)}}\!\!\!\!\!\!\!\!\!\!\!\!\!\hat{\tau}_{s,s+1}^i,\label{Eq.14a}\\
\phi_i(t)&=\epsilon_i\tau_i(t),\label{Eq.14b}\\
q_i&=\frac{T_{w,i}}{T_{d,i}},\label{Eq.14c}
\end{align}
\end{subequations} 
where $(v_k,v_{k+1})\!\in\!{r_{k,d_i}^*\!(t)}$ and $s\neq{k}$ in \eqref{Eq.14a}. The estimated travel times $\tilde{\tau}_{k,k+1}^i$ and $\hat{\tau}_{s,s+1}^i$ are computed using the monitored and estimated traffic flow given in \eqref{Eq.6} and \eqref{Eq.8}, respectively. In addition, $\epsilon_i\!\in\!\{\epsilon_p,\epsilon_a,\epsilon_h\}$, $T_{w,i}\!\in\!\{T_{w,p},T_{w,a},T_{w,h}\}$ and $T_{d,i}\!\in\!\{T_{d,p},T_{d,a},T_{d,h}\}$, relying on the type of vehicle~$i$.

\subsection{Dynamic Trip Equity}
\textit{Dynamic Trip Equity} (DTE) assesses how equitably the transportation resources and services are distributed among all travelers. In a dynamic transportation system, vehicles with overlapping edges in routes compete for road resources, leading to potential congestion. To capture the impact of competing vehicles on traffic conditions and routing decisions, we define the Road Resource Competitors (RRCs) for vehicle $i\!\in\!\mathcal{N}(t)$ at time $t\!\in\![a_k^i,a_{k+1}^i)$ as 
\begin{align}
\mathcal{C}_i(t)\!:=\!\big\{j\!\in\!\mathcal{N}(t)\big| r_{k^{\prime},d_j}^*\!(t)\cap{r_{k,d_i}}\!\neq\!{\emptyset},r_{k,d_i}\!\!\in\!{\mathcal{R}_{k,d_i}}\big\},\!\label{Eq.15}
\end{align}
where $r_{k^{\prime},d_j}^{*}\!(t)$ denotes the optimal route plan of vehicle $j$ at time $t$ to complete the rest of its trip. Vehicles whose routes do not share any overlapping edges with the feasible route of vehicle~$i$ are not directly competing for the same road resources and, thus, are not included in $\mathcal{C}_i(t)$. 

Instead of evaluating the DTE among all travelers in $\mathcal{N}(t)$, which can be computationally inefficient, we propose the DTE defined within the RRCs for each vehicle based on their route plans, ignoring indirect couplings between vehicles. Let $\mathcal{C}_{i,p}(t)$, $\mathcal{C}_{i,a}(t)$, $\mathcal{C}_{i,h}(t)$ be the sets of private, autonomous, and ride-hailing vehicles in $\mathcal{C}_i(t)$, i.e., $\mathcal{C}_i(t)\!=\!\mathcal{C}_{i,p}(t)\cup{\mathcal{C}_{i,a}(t)}\cup{\mathcal{C}_{i,h}(t)}$. We employ the Gini Coefficient~\cite{dorfman1979formula}, an effective measure of inequity in economics and social sciences, to quantify the DTX distribution among travelers. By definition, the DTE for the traveler in vehicle $i$ and those in the RRCs of vehicle $i$ at $t$ is defined as
\begin{align}
\!DTE_i(t)\!:=\!1\!-\!\frac{\sum_{j\in{\mathcal{S}_i(t)}}^{}\!\sum_{j^{\prime}\in{\mathcal{S}_i(t)}}^{}\!\big|DTX_j(t)\!-\!DTX_{j^{\prime}}(t)\big|}{2|\mathcal{S}_i(t)|^2{DTX_{i,\text{mean}}(t)}},\label{Eq.16}
\end{align}
where
\begin{subequations}\label{Eq.17}
\begin{align}
&\mathcal{S}_i(t)=\big(\mathcal{C}_{i,p}(t)\cup\mathcal{C}_{i,a}(t)\big)\uplus \underbrace{\mathcal{C}_{i,h}(t)\dots \uplus \mathcal{C}_{i,h}(t)}_{\text{$m$}},\label{Eq.17a}\\
&{DTX}_{i,\text{mean}}(t)=\frac{1}{|\mathcal{S}_i(t)|}\!\sum_{j\in{\mathcal{S}_i(t)}}^{}\!\!\!\!DTX_j(t).\label{Eq.17b}
\end{align}
\end{subequations}
We assume each ride-hailing vehicle serves $m$ travelers, i.e., each vehicle trip represents $m$ traveler trips. In \eqref{Eq.17a}, $\mathcal{S}_i(t)$ denotes a multiset, representing all traveler trips for vehicles in $\mathcal{C}_i(t)$, with $\uplus$ denoting the multiset sum operation. In \eqref{Eq.17b}, $DTX_{i,\text{mean}}(t)$ is the mean DTX value, where $DTX_j(t)$ is computed via~\eqref{Eq.10}. By definition, $DTE_i(t)\!=\!1$ indicates perfect equity, meaning that $DTX_j(t)$ is identical for all travelers in $\mathcal{S}_i(t)$, whereas $DTE_i(t)\!=\!0$ indicates the maximum inequity.

\subsection{Dynamic Trip Equity in Free-Flow Networks}
This subsection discusses the DTE in a free-flow road network. In scenarios where the infrastructure is sufficient or the number of vehicles in the system is well below the road capacity, congestion is not a concern in route planning. In these cases, vehicles can freely choose the shortest routes for their trips. Let $DTX_{i,0}$ be the DTX for vehicle $i\!\in\!\mathcal{N}(t)$ traversing a free-flow network. Since $\tau_i(t)\!=\!\tau_{\min}^i$, by \eqref{Eq.10}, \eqref{Eq.11b}, and \eqref{Eq.14b}, we have
\begin{align}
DTX_{i,0}=\xi_1\!+\xi_2\frac{\epsilon_{\min}^i}{\epsilon_i}\!+\xi_3\frac{q_{\min}^i}{q_i},\label{Eq.18}
\end{align}
which is independent of time $t$. In line with the DTE presented above, the following result holds.

\begin{mypro}\label{Pro.1}
For any vehicle $i\!\in\!\mathcal{N}(t)$ traveling in a free-flow road network on the trip $(o_i,d_i)$, the best DTE can be achieved between the traveler in vehicle $i$ and those in its RRCs, denoted as $\mathcal{C}_{i,0}(t)$, if it satisfies 
\begin{align}
DTX_{j,0}^p\!=\!DTX_{j,0}^a\!=\!DTX_{j,0}^h,\quad \ \forall{j}\!\in\!\mathcal{C}_{i,0}(t),\label{Eq.19}
\end{align}
where
\begin{subequations}\label{Eq.20}
\begin{align}
DTX_{j,0}^p&=\xi_1\!+\xi_2\frac{\epsilon_{\min}^j}{\epsilon_j}\!+\xi_3,\quad j\!\in\!\mathcal{C}_{i,0}^p(t),\label{Eq.20a}\\
DTX_{j,0}^a&=\xi_1\!+\xi_2\!+\xi_3\frac{q_{\min}^j}{q_j},\quad j\!\in\!\mathcal{C}_{i,0}^a(t),\label{Eq.20b}\\
DTX_{j,0}^h&=\xi_1\!+\xi_2\frac{\epsilon_{\min}^j}{\epsilon_j}\!+\xi_3\frac{q_{\min}^j}{q_j},\quad j\!\in\!\mathcal{C}_{i,0}^h(t).\label{Eq.20c}
\end{align}
\end{subequations}
Here, $\mathcal{C}_{i,0}^p(t)$, $\mathcal{C}_{i,0}^a(t)$, $\mathcal{C}_{i,0}^h(t)$ represent the sets of private, autonomous, and ride-hailing vehicles in $\mathcal{C}_{i,0}(t)$, respectively, with $\mathcal{C}_{i,0}(t)\!=\!\mathcal{C}_{i,0}^p\!\cup\!{\mathcal{C}_{i,0}^a(t)}\!\cup\!{\mathcal{C}_{i,0}^h(t)}$. Similar to \eqref{Eq.15}, for $t\!\in\![a_k^i,a_{k+1}^i)$, $\mathcal{C}_{i,0}(t)$ is of the form
\begin{align}
\mathcal{C}_{i,0}(t)=\big\{j\!\in\!\mathcal{N}(t)\big|\ \underline{r}_{k^{\prime},d_j}\!(t)\cap{\underline{r}_{k,d_i}}\!\neq\!{\emptyset} \big\},\nonumber
\end{align}
where $\underline{r}_{k^{\prime},d_j}\!(t)$ denotes the shortest route for vehicle $j$ to complete its remaining trip and $\underline{r}_{k,d_i}$ is the shortest route from $v_k$ to $d_i$, considering free-flow travel times on edges.
\end{mypro}

\begin{prooflemma}
As indicated by \eqref{Eq.16}, achieving the best DTE requires that $DTE_i(t)\!=\!1$. As $|\mathcal{S}_i(t)|\!\geq\!{1}$ by \eqref{Eq.17a} and $DTX_i(t)\!\in\!(0,1]$ by \eqref{Eq.10}, this condition implies that $DTX_j(t)\!=\!DTX_{j^{\prime}}(t)$, $\forall{j,j^{\prime}}\!\in{\mathcal{S}_i(t)}$. In a free-flow road network, each vehicle $j$ follows the shortest route with the minimum nominal travel time between $(o_j,d_j)$. Thus, $DTX_j(t)\!=\!DTX_{j,0}$ for $j\!\in\!\mathcal{C}_{i,0}(t)$. Moreover, since private vehicles offer the highest convenience and autonomous vehicles typically provide the most economical transportation mode, we have $q_{\min}^j\!=\!q_j$ for $j\!\in\!\mathcal{C}_{i,0}^p(t)$ and $\epsilon_{\min}^j\!=\!\epsilon_j$ for $j\!\in\!\mathcal{C}_{i,0}^a(t)$. Substituting these into \eqref{Eq.18}, the result follows.
\end{prooflemma}

Proposition~1 reveals the conditions for maximizing trip equity in a free-flow road network. These conditions impose constraints on road infrastructure, transportation costs, and vehicle scheduling, which collectively determine the achievable DTX for each vehicle. Note that meeting these conditions ensures that the upper bound of the DTE attainable through route optimization is~$1$. In other words, the D-RGS can improve trip equity only through balancing the DTX via reducing traffic congestion. However, it cannot mitigate inequities caused by other factors, such as travel costs or limitations inherent to certain vehicle types.    
\section{Routing Guidance System Design with Improved DTE}\label{Section IV}   
This section proposes a D-RGS that optimally recommends vehicle routes to enhance $DTE_i(t)$. Let $a_{o_i}^i$ be the departure time of vehicle~$i\!\in\!\mathcal{N}(t)$ from its origin. For vehicle $i$ arriving at a DMP $v_k$ at time $t\!=\!a_k^i$, the D-RGS provides an optimal route recommendation to the vehicle by addressing
\begin{align}
 \max_{r_{k,d_i}\in{\mathcal{R}_{k,d_i}}} DTE_i(a_k^i)\label{Eq.21}
 \end{align}
subject to the constraints
\begin{subequations}\label{Eq.22}
\begin{align}
&\hat{a}_{k+1}^i\!=\!a_k^i\!+\tilde{\tau}_{k,k+1}^i, \ \ (v_k,\!v_{k+1})\!\in\!{r_{k,d_i}},\!\!\label{Eq.22a}\\
&\hat{a}_{s+1}^i\!=\!\hat{a}_s^i\!+\hat{\tau}_{s,s+1}^i, \ \ (v_s,\!v_{s+1})\!\in\!{r_{k,d_i}}, \ s\!\neq\!{k},\label{Eq.22b}\\
&a_{k,k+1}^j\!=\!a_k^j,\ \ j\!\in\!\mathcal{C}_i(a_k^i), \ j\!\neq\!{i},\label{Eq.22c}\\
&\hat{a}_{h,h+1}^j\!=\!\hat{a}_h^j,\ j\!\in\!\mathcal{C}_i(a_k^i), \ j\!\neq\!{i},\ h\!\neq\!{k},\label{Eq.22d}\\
&\hat{a}_{h+1}^j\!=\!\hat{a}_h^j\!+\!\tau^0_{h,h+1}, \ \ (v_h,\!v_{h+1})\!\in\!{r_{k^{\prime},d_j}^*\!(a_k^i)}, \label{Eq.22e}\\
&\ \ \ \ \ \ \ \ \ \ \ \ \ \ \ \ \ \ \ \ \ \ \ \ \ \  j\!\in\!\mathcal{C}_i(a_k^i),\ j\!\neq\!{i},\nonumber\\
&\eqref{Eq.3}\!-\!\eqref{Eq.6},\eqref{Eq.8}\!-\!\eqref{Eq.17}.\nonumber
 \end{align} 
 \end{subequations}
 
In the optimization problem formulated above, the target is to maximize $DTE_i(t)$, as defined in \eqref{Eq.15}-\eqref{Eq.17}, by optimally selecting a route $r_{k,d_i}$ from the set of all feasible routes $\mathcal{R}_{k,d_i}$. Constraints \eqref{Eq.22a} and \eqref{Eq.22b} provide estimated dynamics for vehicle~$i$, while \eqref{Eq.22c}-\eqref{Eq.22e} describe the estimated dynamics for vehicles $j\!\in\!\mathcal{C}_i(a_k^i)$, $j\!\neq\!{i}$. Travel times $\tilde{\tau}_{k,k+1}^i$ and $\hat{\tau}_{s,s+1}^i$, derived from the monitored and estimated traffic flow, are computed by \eqref{Eq.3}-\eqref{Eq.6}, \eqref{Eq.8} and \eqref{Eq.9}. The computation of the DTX is based on the constraints \eqref{Eq.10}-\eqref{Eq.14}. 

The problem \eqref{Eq.21} subject to \eqref{Eq.22} can be addressed by dynamic programming~\cite{bertsekas2019reinforcement,10209062} each time a vehicle reaches a DMP, and the optimal solution is denoted as $r_{k,d_i}^*$. Note that, if vehicle $i$ has no RRCs at $v_k$, i.e., $|\mathcal{C}_i(a_k^i)|\!=\!1$, the D-RGS provides the shortest route for it, as $DTE_i(a_k^i)\!=\!1$ naturally holds in this case. Unlike one-time optimization, the D-RGS continuously optimizes and updates the route recommendation at each MDP by maximizing the dynamic trip equity $DTE_i(a_k^i)$. This facilitates incorporating real-time traffic data and coping with environmental uncertainties.
\begin{myrem}\label{Remark2}
The computational complexity for solving \eqref{Eq.21} subject to \eqref{Eq.22} is determined by $|\mathcal{R}_{k,d_i}|$ and $|\mathcal{C}_i(a_k^i)|$. To achieve a trade-off between the control performance and the computational efficiency, Yen's algorithm~\cite{martins2003new} is employed to construct $\mathcal{R}_{k,d_i}$ at each DMP, which selects the $L$ shortest routes from $v_k$ to $d_i$ instead of using all feasible routes. Notice that both $|\mathcal{R}_{k,d_i}|$ and $|\mathcal{C}_i(a_k^i)|$ decrease as the vehicle approaches its destination, which further enhances the computational efficiency of our method. 
\end{myrem}
\begin{myrem}\label{Remark3}
The D-RGS provides socially optimal route recommendations to individual vehicles that enhance trip equity across travelers. In practice, vehicles may deviate from the recommended routes based on their own optimization metric or preference~\cite{Xu2025deviation}, such as minimizing travel time or selecting routes with better road conditions. Our follow-up work~\cite{Li2025arouting} focuses on effective incentive design to improve route compliance and social optimality.
\end{myrem}

\section{Simulation Studies}\label{Section V}
We consider an urban road network in Boston comprising $45$ road intersections, $8$ origin nodes, and $5$ destination nodes. The simulation involves $1,000$ vehicles with random departure times between $08\!\!:\!\!00$-$10\!\!:\!\!00$, including $500$ private vehicles, $300$ autonomous vehicles, and $200$ ride-hailing vehicles, where each ride-hailing vehicle serves two travelers (i.e., $m\!=\!2$ in \eqref{Eq.17a}). The OD pairs are randomly selected from the origin and destination sets. We assume all vehicles travel at a free-flow speed of $27$ km/h, and the nominal travel time on each edge is obtained from \emph{OpenStreetMap}~\cite{OpenStreetMap}. The maximum traffic flow is set as $5$ vehicles per minute per lane. The parameter settings for each vehicle type, satisfying the conditions in Proposition~\ref{Pro.1}, are provided in Table~\ref{Table1}. To enhance computational efficiency, Yen's algorithm is employed to generate $7$ shortest routes from a DMP $v_k$ to the destination $d_i$ when forming the feasible route set $\mathcal{R}_{k,d_i}$.

\begin{table}[t]
\caption{Parameter Settings of Each Vehicle Type} 
\vspace{-5pt}
\centering 
\renewcommand{\arraystretch}{1.5} 
\begin{tabular}{c|c|c|c|c|c|c}
  \hline\hline
  Vehicle type & $\xi_1$ & $\xi_2$ & $\xi_3$ & $\epsilon_i$[\textdollar/min] & $T_{w,i}$[min] & $T_{d,i}$[h] \\  
  \hline
  $i\!\in\!\mathcal{N}_p(t)$ & $0.4$ & $0.4$ & $0.2$ & $0.27$ & $2$ & $24$ \\  
  \hline
  $i\!\in\!\mathcal{N}_a(t)$ & $0.4$ & $0.4$ & $0.2$ & $0.1485$ & $15$ & $18$ \\  
  \hline
  $i\!\in\!\mathcal{N}_h(t)$ & $0.4$ & $0.4$ & $0.2$ & $0.1536$ & $6$ & $12$ \\  
  \hline\hline
\end{tabular}
\label{Table1}
\end{table}

\begin{figure}[t]
     \centering
     \includegraphics[width=0.9\linewidth]{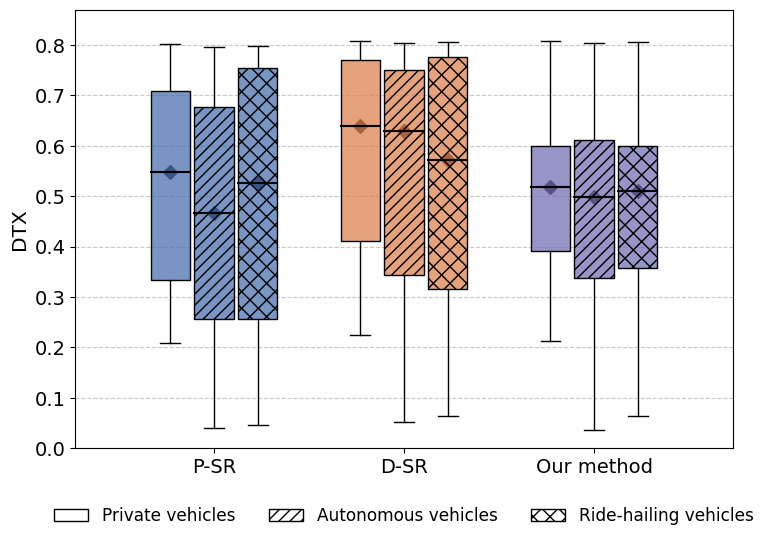}
     \vspace{-5pt}
      \caption{Comparison of the DTX. The results show that DTXs in our method cluster more closely around the average value.}
      \label{Fig.1}
   \end{figure}
   
\subsection{Results and Analysis}
We compare the proposed method with the pre-planned and dynamic shortest-route strategies, where the pre-planned shortest-route (P-SR) strategy minimizes the trip time based on the nominal travel time, while the dynamic shortest-route (D-SR) strategy incorporates the monitored and estimated traffic flow information to optimize the trip time dynamically. Traffic congestion is monitored in real-time at intervals of $\Delta{t}\!=\!60$ seconds. The DTX of each vehicle, integrating the actual travel time in \eqref{Eq.10} under the three strategies, is shown in Fig.~\ref{Fig.1}. The DTEs for all completed trips in P-SR, D-SR, and our methods are $0.734$, $0.777$, and $0.818$, respectively. As we observe, the P-SR strategy results in a wide spread of DTX values, indicating large inequity across individual trips. The D-SR strategy improves performance by increasing the median DTX, but still exhibits considerable variability, with a substantial number of vehicles experiencing low trip equity. In contrast, our proposed method achieves a more centralized distribution of DTX across all vehicle types, showing a more consistent and balanced trip equity.
\begin{figure}[t]
     \centering
     \includegraphics[width=0.85\linewidth]{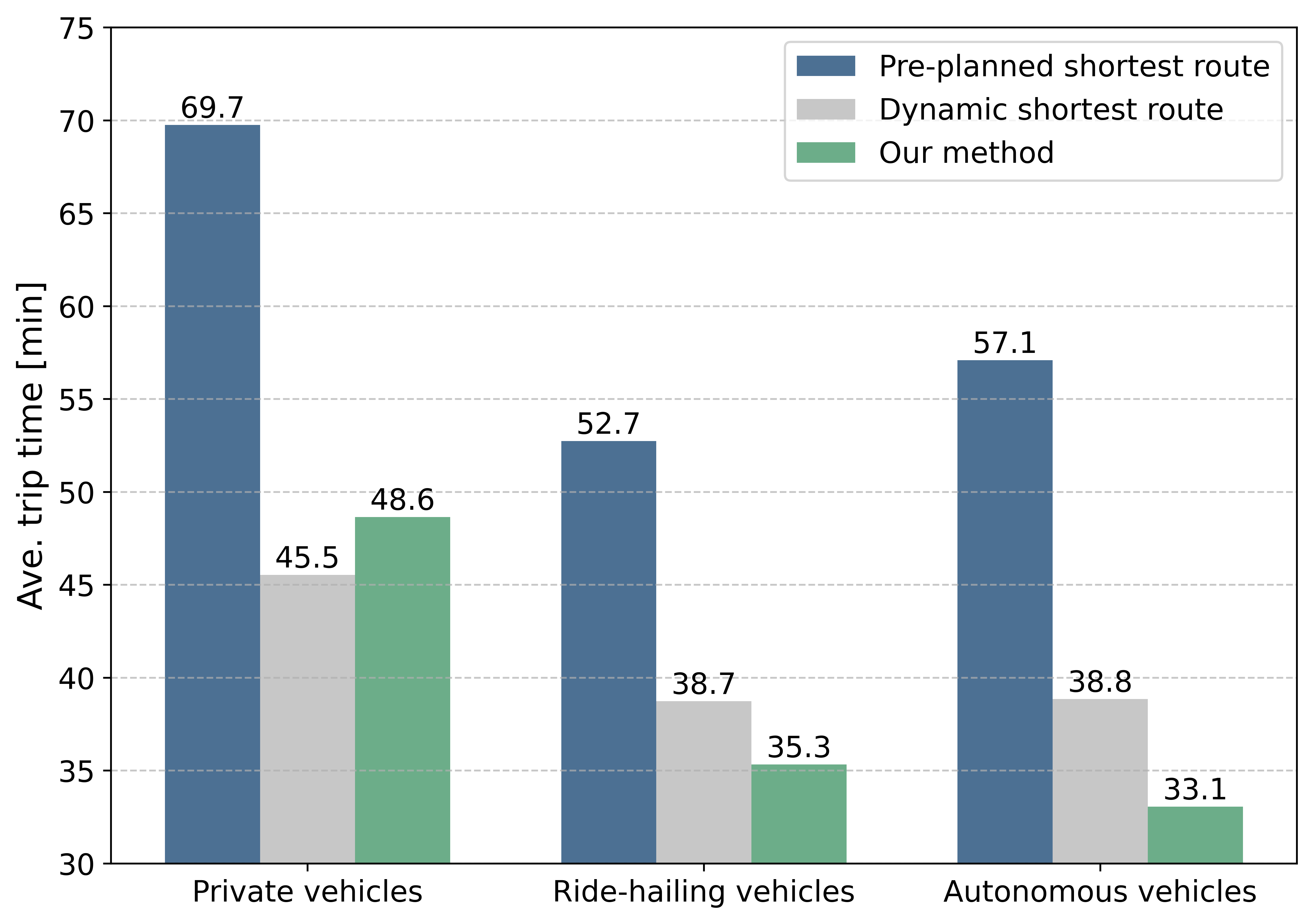}
     \vspace{-5pt}
      \caption{Comparison of the average trip time.}
      \label{Fig.2}
   \end{figure}
   
Fig.~\ref{Fig.2} compares the average trip time in the three strategies. As is shown, the proposed method significantly reduces the average trip time for private, ride-hailing, and autonomous vehicles by approximately $30\%$, $33\%$, and $42\%$, respectively, leading to per-trip savings of about $\text{\$}5.7$, $\text{\$}2.7$, and $\text{\$}3.6$ for each vehicle type. Compared to the D-SR strategy, our method increases the average trip time for private vehicles by $3.1$ minutes, while reducing trip times for both ride-hailing and autonomous vehicles by approximately $8.8\%$ and $14.7\%$, respectively. This outcome aligns with real-world expectations, where private vehicle owners typically belong to higher-income groups, and a slight increase in their travel time as well as travel costs leads to equity enhancement for other travelers. These results demonstrate that our method effectively redistributes travel costs across vehicle types through route optimization, contributing to a more equitable and efficient transportation system. 

\section{Conclusion}\label{Section VI}
In this paper, we presented a framework to quantify trip quality and equity in a dynamic emerging transportation system, integrating trip time, cost, and convenience. We developed a routing guidance system to optimize vehicle route recommendations, incorporating real-time and anticipated traffic congestion while improving trip equity. In addition, we established conditions that ensured perfect trip equity in free-flow networks. Our simulation studies on a Boston road network demonstrated that the proposed approach increased trip equity from $0.734$ to $0.818$ compared to the pre-planned shortest-route strategy, leading to $11.4\%$ improvement and a more equitable transportation system. Future work will focus on learning the compliance rate and designing effective incentive mechanisms to enhance drivers' alignment with routing recommendations.
\bibliographystyle{IEEEtran}
\bibliography{CDC_2025,IDS_2025}
\end{document}